\documentclass[%
 reprint,
superscriptaddress,
 amsmath,amssymb,
 aps,
]{revtex4-2}

\usepackage{xcolor}
\usepackage{graphicx}
\usepackage{dcolumn}
\usepackage{bm}
\usepackage{subfigure}

\begin{document}

\preprint{APS/123-QED}

\title{Chaotic dynamics creates and destroys branched flow}

\author{Alexandre Wagemakers}
\affiliation{Nonlinear Dynamics, Chaos and Complex Systems Group, Departamento de  F\'isica, Universidad Rey Juan Carlos,
Tulip\'an s/n, 28933 M\'ostoles, Madrid, Spain}
\email{alexandre.wagemakers@urjc.es}

\author{Aleksi Hartikainen}
\affiliation{Computational Physics Laboratory, Tampere University, Tampere 33720, Finland}
\email{aleksi.hartikainen@tuni.fi}

\author{Alvar Daza}
\affiliation{Nonlinear Dynamics, Chaos and Complex Systems Group, Departamento de  F\'isica, Universidad Rey Juan Carlos,
Tulip\'an s/n, 28933 M\'ostoles, Madrid, Spain}
\affiliation {Department of Physics, Harvard University, Cambridge, Massachusetts 02138, USA}
\email{alvar.daza@urjc.es}

\author{Esa R\"as\"anen}
\affiliation{Computational Physics Laboratory, Tampere University, Tampere 33720, Finland}
\affiliation {Department of Physics, Harvard University, Cambridge, Massachusetts 02138, USA}
\email{esa.rasanen@tuni.fi}

\author{Miguel A.F. Sanju\'{a}n}
\affiliation{Nonlinear Dynamics, Chaos and Complex Systems Group, Departamento de  F\'isica, Universidad Rey Juan Carlos,
Tulip\'an s/n, 28933 M\'ostoles, Madrid, Spain}
\email{miguel.sanjuan@urjc.es}


\date{\today}

\begin{abstract}
The phenomenon of branched flow, visualized as a chaotic arborescent pattern of propagating particles, waves, or rays, has been identified in disparate physical systems ranging from electrons to tsunamis, with periodic systems only recently being added to this list. Here, we explore the laws governing the evolution of the branches in periodic potentials. On one hand, we observe that branch formation follows a similar pattern in all non-integrable potentials, no matter whether the potentials are  periodic or completely irregular. Chaotic dynamics ultimately drives the birth of the branches. On the other hand, our results reveal that for periodic potentials the decay of the branches exhibits new characteristics due to the presence of infinitely stable branches known as superwires. Again, the interplay between branched flow and superwires is deeply connected to Hamiltonian chaos. In this work, we explore the relationships between the laws of branched flow and the structures of phase space, providing extensive numerical and theoretical arguments to support our findings.
\end{abstract}

\maketitle


\section{\label{sec:intro}Introduction}

Electrons, lasers, tsunamis, and ants have at least one thing in common: they all display branched flow~\cite{heller2021branched, topinka2001coherent, patsyk2020observation, degueldre2016random, mok2023branched}. Whenever a wave propagates through a weakly refracting medium, flow is expected to accumulate along certain directions, forming structures called branches. Among the different examples, tsunamis serve as the most dramatic illustration of the powerful implications of branched flow. Branches can carry an unusual amount of energy towards unpredictable points in space.

Despite being a ubiquitous phenomenon with remarkable effects, basic knowledge of the mechanisms behind branched flow is still lacking. It has been well established that branched flow occurs in the semiclassical limit~\cite{heller2018semiclassical}. This means that the wavelength should be small compared to the characteristic length of the refracting medium. Deflections must be small too, so that branched flow is mostly made of {\em forward} scattering. However, the precise relations between the flow patterns and the shape of the potentials producing them still remains unclear. Indeed, almost all the literature on branched flow focuses solely on the so-called random potentials. It is worth mentioning that the term random potential might be confusing, since branched flow is a completely deterministic process. However, given its wide use in the existing literature, here we refer to random potentials when talking about smooth potentials lacking any spatial order. In all the examples cited so far, waves evolve in these smooth landscapes with no perceptible spatial pattern. 

Spatial disorder of the potential is not a requirement for the dynamics of branched flow. In fact, periodic potentials have recently been identified to produce branched flow, opening up a whole new realm of possibilities~\cite{daza2021propagation}. From superlattices such as twisted bilayer graphene~\cite{cao2018unconventional} to photonic crystals~\cite{joannopoulos1997photonic}, branched flow could be observed if the right conditions are met. Furthermore, in periodic potentials, some branches remain indefinitely stable, forming so-called superwires.

\begin{figure*}
    \centering
    \includegraphics[width = \textwidth]{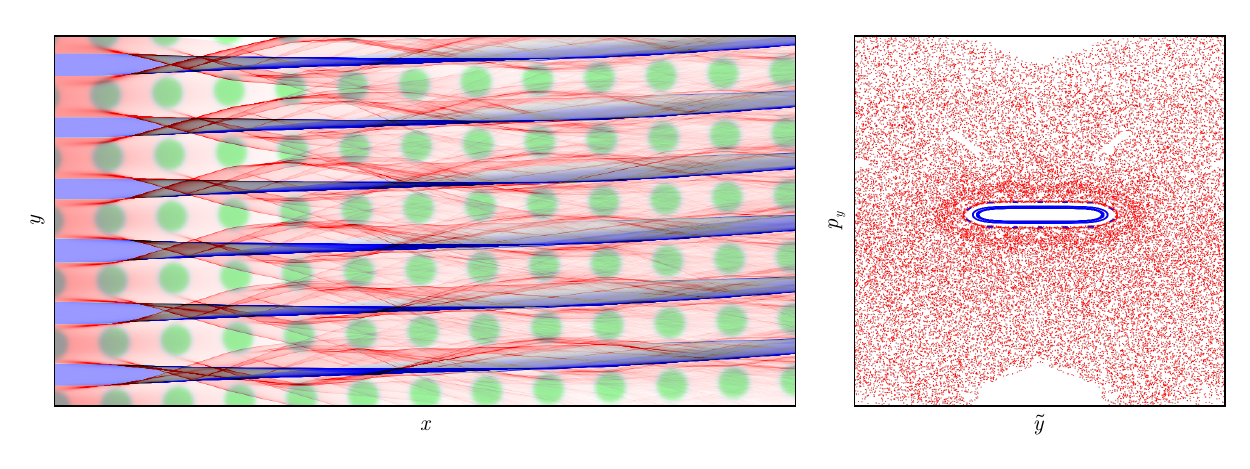}
    \caption{Branched flow in the Fermi periodic potential of Eq.~(\ref{eq:FermiV}). In the left panel, red tones indicate branched flow related to chaotic trajectories (positive maximum Lyapunov exponent), whereas blue tones are used for superwires (zero maximum Lyapunov exponent). The right panel depicts the Poincaré section along the direction transverse to the potential $\tilde{y}$ using the same color coding. Superwires are caused by KAM stable islands (in blue) while branched flow is related to the chaotic sea (in red).}
    \label{fig:BF_Fermi}
\end{figure*}

Fig.~\ref{fig:BF_Fermi} shows branched flow (in red) and superwires (in blue) as the result of a plane wave evolving in a periodic potential. In the superwires, it is the dynamics that prevent the spreading of the flow, unlike being energetically trapped as in a waveguide. In fact, superwires are related to stable dynamics, while branched flow is related to chaotic dynamics, as can be observed in the phase space portrait of Fig.~\ref{fig:BF_Fermi}. Potential applications of superwires include superconductivity and beam focusing.

Early works investigated the laws of branched flow only in random potentials~\cite{kaplan2002statistics, metzger2010universal}. Therefore, it is natural to ask how these laws translate in the case of a periodic potential. Our goal in this work is to understand how branches emerge and disappear in periodic potentials. A simple visual inspection might indicate that branched flow looks similar at both random and periodic potentials, but the presence of superwires implies important differences in the dynamics that require detailed investigation. In this paper, we discuss how prior knowledge on branched flow needs to be revised for periodic media.

The manuscript is organized as follows. First, the dynamics of branched flow and the numerical techniques used to characterize it are described in Sec.~\ref{sec:methods}. Our main findings are included in Sec.~\ref{sec:results}, where we study the laws of the birth and death of the branches in periodic potentials. Finally, we discuss the implications and perspectives of our work in Sec.~\ref{sec:discussion}.

\section{\label{sec:methods}Methods}
Here, we detail the numerical and theoretical tools employed along the manuscript.

\subsection{\label{sec:semiclassic}Semiclassical dynamics of branched flow}
There are two conditions that make branched flow a semiclassical phenomenon~\cite{heller2018semiclassical}. First, the wavelength must be smaller than the characteristic length of the potential. Second, the potential height must be sufficiently small, allowing the wave to be gently deflected without backscattering:

\begin{align}
    \lambda &\ll l_{pot} 
      \\V_{max}  &\ll E_{wave}.
\end{align}

In this regime, wave propagation can be replaced by ray-tracing of the appropriate manifolds. For example, a plane wave can be described as a set of parallel rays with fixed energy, or a point source can be modeled using rays coming out from a single point in space in every possible direction.

The small deflection condition still enables a further simplification. It is possible to neglect longitudinal dynamics and concentrate only in the transverse direction. In this work, we focus on two-dimensional branched flow. Therefore, by ignoring the longitudinal direction, we are reducing the problem to just one dimension. This is the quasi-2D approximation \cite{kaplan2002statistics}, where the motion associated to the transverse ($y$) and longitudinal ($x$) directions is described by

\begin{align}
    &\dot{p_y}  =  -\frac{\partial V}{\partial {y}} 
   \\ &\dot{y}  = p_y
      \\ &\dot{p_x} =  0\\
      &\dot{x} = p_x = k.
\end{align}
Here $k$ is a constant that can be set equal to one without loss of generality and $V(x,y)$ represents the two-dimensional potential.  By construction, there is no backscattering in this approximation. Notice that quasi-2D dynamics in a weak Hamiltonian potential is similar to the paraxial approximation in optics considering the wave propagation along one direction with small transverse variations.

Given the trivial dynamics along the longitudinal axis, we can reduce the model even further and get the a discrete map for the transverse dynamics
 \begin{align}
    p_y(n+1) & =  p_y(n) -\left(\frac{\partial V}{\partial y}\right)_{(x,y)=(x_n,y_n)}\\
    y(n+1) & = y(n) + p_y(n+1).
    \label{eq:knd}
 \end{align}

This area-preserving map is usually referred to as the kick and drift map~\cite{heller2021branched}. 

\begin{figure*}[!t]
    \centering
    \includegraphics[width=\textwidth]{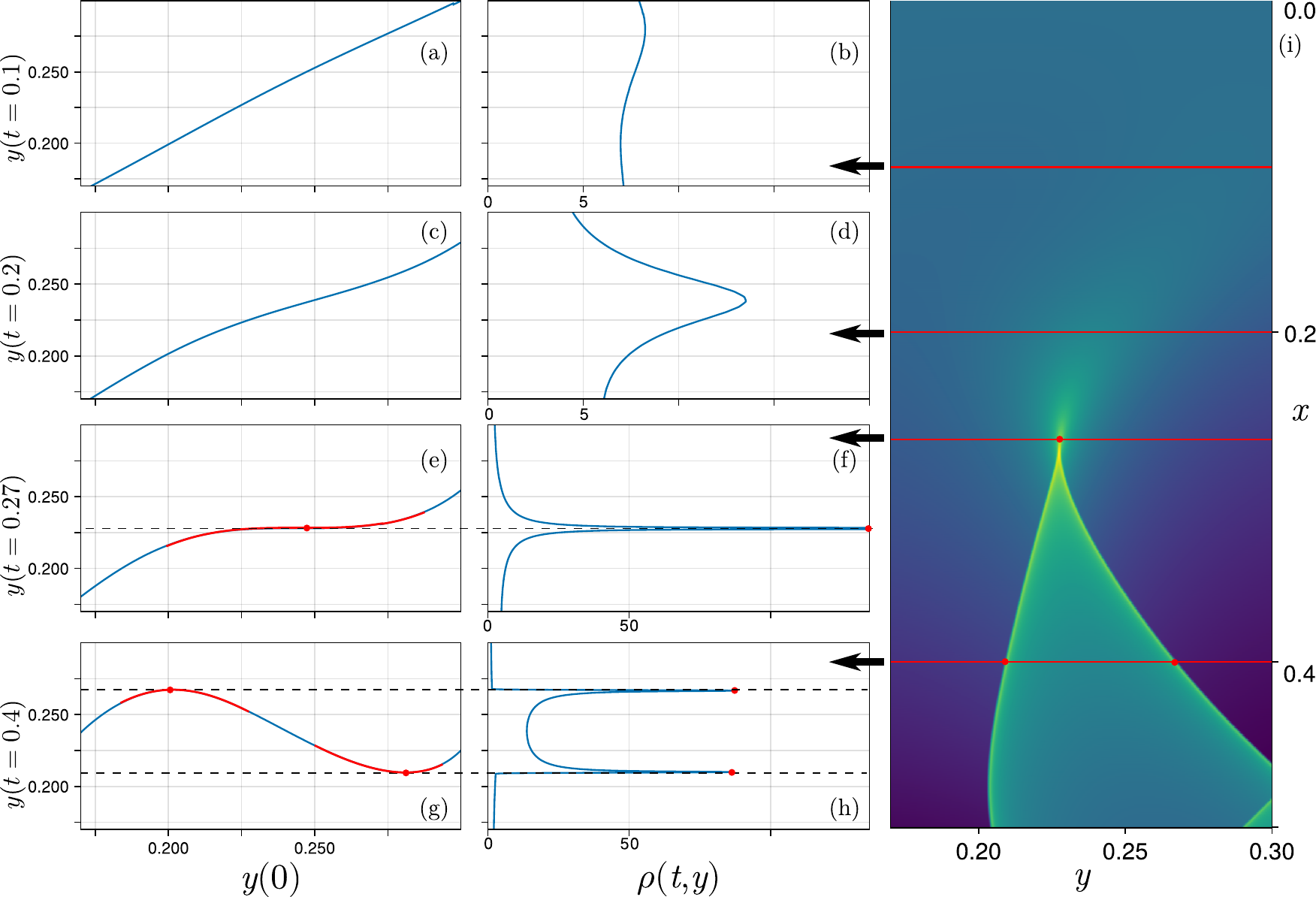}
    \caption{Visible branch detection with the manifold tracking algorithm. Trajectories in a random correlated potential with correlation length $l_c = 0.1$ are launched from a planar front wave represented in the right panel (i). Rays propagate downward. Panels in the left column are plots of the position $y(t)$ of the rays at time $t$ ordered according to their starting position $y(0)$. Panels in the central column represent an estimation of the rays density. The horizontal axis $\rho(t,y)$ holds for the density and the vertical axis is the coordinate $y(t)$. The representation has been tilted 90$º$ to match the manifold coordinates. The figure has been divided into 4 time frames to explain the cusps and branches formation. (a)-(b) At time $t = 0.1$ the manifold is nearly flat and the density of rays is almost uniform. (c)-(d) As the wavefront begins to propagate a time $t=0.2$, the shape of the manifold in panel (c) begins to wiggle as rays move influenced by the potential. The density begins to peak (panel (d)) although it is not a visible branch. (e)-(f) At time $t=0.27$ the cusp is formed. Its manifestation in the manifold is a null derivative marked with a red dot in (e) and a peak in ray density in (f). The part of the manifold painted in red corresponds to $|\partial y(t)/\partial y(0)| < 1$. Along with the zero derivative, it is our criterion to detect visible branches in the physical space. (g)-(h) The last time frame $t=0.4$ shows the two branches born from the cusp. The rays' density in (h) perfectly corresponds with the red portion of the manifold in (g). The two branches are clearly visible in panel (i).
    }
    \label{fig:branch_detect}
\end{figure*}

\subsection{\label{sec:counting}Counting branches}
Intuitively, branches could be defined as unusual accumulation of flow in certain areas of space. Nevertheless, it can be hard to find a precise way to quantify that intuition. Among the different measures proposed in the literature, we have chosen an improved version of the manifold tracking algorithm~\cite{metzger2010branched} because it requires the smallest number of arbitrary parameters to be adjusted and produces the most robust results. Next, we explain its functioning.

The manifold tracking algorithm comprises two stages. For each time step, the algorithm looks at the evolution of the transverse coordinate $y(t)$ as a function of the corresponding initial condition $y(0)$, as shown in the left column of Fig.~\ref{fig:branch_detect}. Manifolds evolving in a weak potential may undergo cusp catastrophes, which suppose an accumulation of rays in a single point. The density of the rays $\rho(t,y)$ shows an acute peak at the cusps, as shown in panel (f) of Fig.~\ref{fig:branch_detect}. These cusps could be viewed as the origin of the branches, i.e, the seeds of the branched flow. In two dimensions, a cusp results in two caustics, or focal lines, corresponding to two peaks in the density $\rho(t,y)$, as depicted in Fig.~\ref{fig:branch_detect}-(h). In free-space propagation, cusps exhibits a Pearcey diffraction pattern~\cite{Berry2020}. However, because of the non-integrability of the potential the pattern gets smoothed, resulting in two localized branches visible in the plot. The following condition is met for the caustics

 \begin{equation}
        \left(\frac{\partial y(t)}{\partial y(0)}\right)_{y=y_\text{caustic}} = 0.
        \label{eq:cond1}
\end{equation}

Hence, we examine the local maxima and minima in the plot of $y(t)$ versus $y(0)$ to determine the number of branches, marking the initial phase of the manifold tracking algorithm. The underlying concept is that the ray density $\rho(t,y)$ corresponds to the projection of the $y(t)-y(0)$ plot onto the vertical axis, with extremes yielding density peaks. We can see that for the cusps, besides Eq.~\ref{eq:cond1}, the following condition also applies
 \begin{equation}
        \left(\frac{\partial^2 y(t)}{\partial y(0)^2}\right)_{y=y_\text{cusp}} = 0.
\end{equation}
 
The process of cusp formation continues indefinitely in the ray-tracing approximation; however, beyond a certain threshold, new cusps cease to generate visible branches. When the ray density accumulated by these cusps falls below the background density, they are engulfed by the diffusion process. Consequently, relying solely on cusp formation to identify branches becomes insufficient, necessitating additional criteria to account for the branches.

The second stage of the manifold tracking algorithm relies on the fact that the rays concentrate in a small region around the caustics. More precisely, we define a visible branch as a continuous piece of the manifold that has stretched less than a certain amount: 
\begin{equation*}
\left|\frac{\partial y(t)}{\partial y(0)}\right| \equiv |s_{y(0)}(t)|  < \varepsilon, 
\end{equation*} 
where $s_{y(0)}(t)$ is the stretching experienced by a small piece of the manifold initially at $y(0)$ after a time $t$. 
 
If the slope of a continuous piece of the manifold is small enough and contains a minimum of $N_r$ rays, then the branch becomes visible. 
 
Figure~\ref{fig:branch_detect} illustrates the whole detection process. Panels (a), (c), (e), and (g) show the transverse position $y(t)$ of the particle at four different times $t$ as a function of their initial position $y(0)$. The caustics are located at the stationary points with null derivative and marked with red dots. The pieces of the manifold highlighted in red are the regions with $|s_{y(0)}(t)| < 1$. The pieces of manifold in red are the visible branches as long as they contain at least $N_r$ rays. In Fig.~\ref{fig:branch_detect} (b), (d), (f) and (h), we represent the ray density $\rho(y,t)$ as a function of the $y$ axis. The dashed line corresponds to the position of the branches. Finally, panel Fig.~\ref{fig:branch_detect}-(i) displays the ray density in the physical space $x-y$. The branch detection depends now on two parameters: the threshold $\varepsilon$ for the stretching of the manifold and the minimal number $N_r$ of rays contained in the piece of the manifold. The number of branches detected is stable for reasonable variations of these parameters. We have chosen $\varepsilon = 1$ and $N_r= 3$ for the rest of the study.

\subsection{\label{sec:potentials}Potentials}

As already mentioned in the Introduction, the main goal of this paper is to investigate the laws governing the birth and death of the branches in periodic potentials. Instead of focusing in a particular physical system or model, we aim to find general results applicable to any two-dimensional periodic system exhibiting branched flow. Therefore, the potentials here play the role of mere test subjects that allow us to explore the behavior of the branches.

Perhaps the most straightforward two-dimensional periodic potential that comes to mind is 
\begin{equation}\label{eq:sinV}
        V(x,y) = \sin x + \sin y.
\end{equation}
However, as discussed in~\cite{daza2021propagation}, this integrable potential cannot produce branched flow. Nevertheless, this sine potential serves as a reference to compare and study the particular features of branched flow, we will refer to it as periodic integrable potential.

A more interesting periodic potential can be built through the so-called Fermi potential~\cite{lorentzgas}, which can be written as
\begin{equation}\label{eq:FermiV}
        V(\vec{r})=\sum_{j=1}^N V_0/\left[1+\exp(|\vec{r}-\vec{r_{0j}}|/\sigma)\right],
\end{equation}
where the parameters $V_0$ and $\sigma$ determine the height and steepness of the potential respectively, and $r_{0j}$ represents the positions in a square lattice (although other periodic structures could be considered). This potential is non-integrable and previous works have already used it to investigate branched flow and other diffusive phenomena~\cite{daza2021propagation, lorentzgas}. In this case, chaotic and periodic dynamics are present in phase space depending on the initial conditions.  

Finally, a random Gaussian correlated potential with correlation length $l_c$ is also used as a benchmark: 
\begin{equation}
    \langle V(\vec r) \rangle = 0 
\end{equation}
\begin{equation}
\langle{V(\vec {r_1}) V(\vec {r_2})}\rangle = c(|\vec{r_1} - \vec{r_2}|) .
\label{eq:randomV}
\end{equation}
Here $c$ denotes a correlation function $c(r) = v_0^2 \exp(-r^2/l_c)$. The branch statistics for this random potential have been well established~\cite{kaplan2002statistics, metzger2010universal} and provide a reference point for our research. 

\section{\label{sec:results}Results}

Now, we present the main numerical and theoretical results concerning the laws of the birth and death of the branches in a periodic potential.

\subsection{\label{sec:birth}Birth of the branches}
As a wave or bundle of rays propagates in a weak potential, cusps arise and branches follow. Thus, we must account for the apparition of cusps to understand the creation of the branches. In~\cite{kaplan2002statistics}, it is argued that cusps grow exponentially with time in \textit{random} potentials,
\begin{equation}
    N \propto e^{\alpha t},
\end{equation}
where the parameter $\alpha$ is related to the Lyapunov exponent and the inhomogeneity of the stretching. 

We can try to understand this exponential growth in the context of \textit{periodic} potentials by means of a simplified argument. Looking at Fig.~\ref{fig:branch_detect}, we can see how the weak potential concentrates the rays in the cusp, and then two branches emerge as a maximum and minimum in the $y(t)$ versus $y(0)$ plot. The total effect is that the set of initial conditions $y(0)$ get transformed into something that resembles a sine function $ \sin(y(0))$. Given the periodicity of the potential, we can loosely assume that this action is repeated indefinitely, so that $y_{n+1}=\mu \sin (y_n)$, where the index $n$ refers to the unit cell of the periodic potential and $\mu$ to the characteristics of the potential. This one-dimensional map folds the initial conditions $y_0$ at each step and, for sufficiently large $\mu$, the number of folds increases exponentially with $n$. In other words, the number of branches increases exponentially with time, since the local maxima and minima of $y_n$ versus $y_0$ correspond with the branches, as explained in Sec.~\ref{sec:counting}. This argument captures the exponential growth of the branches, although it fails in some aspects. In particular, there is a degeneracy in the location of the branches, since in the sine map all the maxima/minima have the same value of $y$, meaning that all the branches are concentrated in a single point. This happens because the dynamics is actually more complicated than what a one-dimensional map can capture. At least a two-dimensional map such as the kick and drift map of Eq.~\ref{eq:knd} is needed to faithfully reproduce the main features of branched flow. The dynamics of the kick and drift map stretches and folds the initial manifold in phase space producing several vertical segments after some time. At these points, we have that $\left(\frac{\partial p_y}{\partial y}\right)=\infty$, which is the cusp condition. 

If the dynamics is periodic, the cusps will get undone and then reappear again in the same positions after some time. Thus, for periodic motion, the number of cusps remains constant on average. An example of periodic integrable potential has been explicitly solved both theoretically~\cite{BerryODell1999} and experimentally~\cite{LucasBiquard1932}. In this scenario, cusps appear periodically along the propagation axis. However, for chaotic dynamics, the cusps are formed at new unpredictable positions produced by the kick and drift evolution rule. This process continues as time goes by, creating a fractal structure which is ultimately responsible for the exponential growth of the cusps.

\begin{figure}
    \centering
    \includegraphics[width=0.5\textwidth]{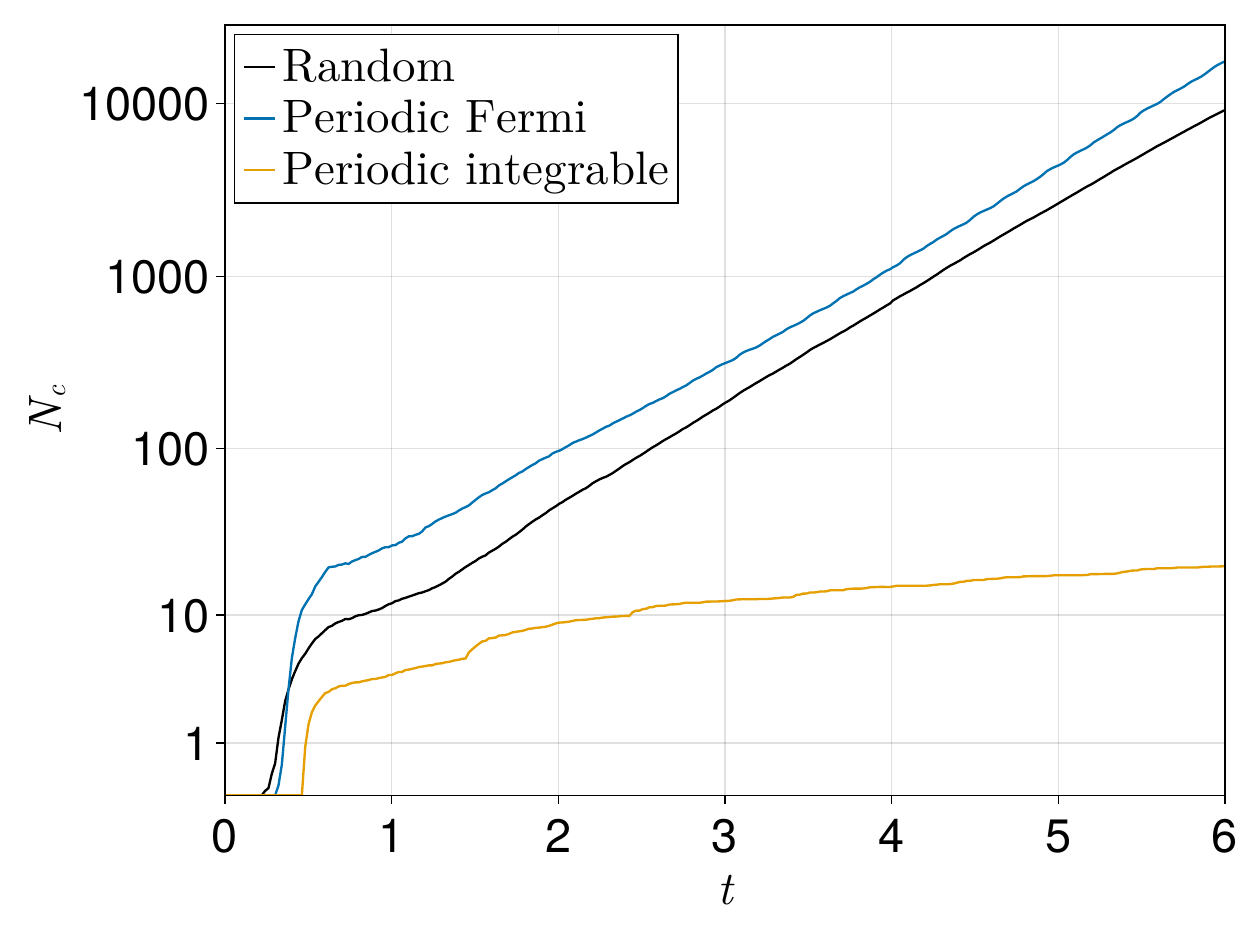}
    \caption{Number of caustics over time for various potentials. Both the random potential described by Eq.~\ref{eq:randomV} and the periodic Fermi potential defined by Eq.~\ref{eq:FermiV} exhibit comparable exponential growth patterns. Conversely, the integrable potential displays distinct behavior, appearing to reach an upper limit. Initially, the number of visible branches $N_b$ equals the number of caustics $N_c$. However, this correlation diminishes over time since the formed branches do not contain enough rays. The caustics are detected with the manifold tracking algorithm \cite{metzger2010branched}. The time scale stops a $t=6$ due to a numerical limitation: the exponential growth cannot be maintained with a finite number of initial rays.}
\label{fig:ExpBirth}
\end{figure}

We conducted numerical computations to observe the evolution of caustics over time using various potentials, aiming to validate our theoretical insights. We can observe in Fig.~\ref{fig:ExpBirth} that the number of caustics $N_c$ exhibits exponential growth for random potentials, as anticipated. Similarly, for non-integrable periodic potentials like the Fermi potential of Eq.~(\ref{eq:FermiV}), exponential growth is evident. However, for integrable periodic potentials, growth is constrained by an upper limit. This underscores that spatial randomness in the potential is not a prerequisite for branched flow. Chaotic dynamics is the necessary ingredient for branched flow, achievable with both periodic and random potentials.

The number of caustics $N_c$ is an excellent proxy for the number of branches $N_b$ up to a dozen correlation length or so for the first periods. A branch is a caustic with a sufficient density of rays to become visible, we include this notion to count correctly $N_b$ in the following section.

\subsection{\label{sec:death}Death of the branches}
Branched flow is a transient regime that precedes homogeneous diffusion. Sufficiently far from their origin, branches fade away and eventually die. Adopting a semiclassical perspective and thinking in terms of phase space, we can understand this effect as follows. The initial manifold gets stretched and folded creating the cusps that we studied in the previous section. These cusps give birth to the branches, which continue to live as long as the dynamics do not smear them too much. For \textit{random} potentials, this means that the number of visible branches grows exponentially, peaks, and then decays exponentially with time~\cite{kaplan2002statistics},
\begin{equation}\label{eq:expdie}
    N_b \propto e^{-\Omega t},
\end{equation}
where $\Omega$ is also related to the stretching factor distribution of the transverse direction. We must recall that in the ray-tracing approximation, the number of caustics keeps increasing indefinitely, since the dynamics keep stretching and twisting the initial manifold and the fractal structure of cusps goes on at all scales. However, as the whirls and curls diminish in size, the caustics eventually merge with the background density, becoming indistinguishable from it. Furthermore, if we want to make the connection with the quantum mechanics counterpart, the phase space has some finite resolution given by $\hbar$, blurring any details finer than this scale. Therefore, after the initial rise, cusps and visible branches become uncorrelated. Although for long times the number of caustics is no longer useful, we can still count branches directly from the data using the manifold tracking algorithm explained in Sec.~\ref{sec:counting}. 

In non-integrable periodic potentials, an exponential decay is also to be expected, given that the chaotic dynamics is similar to the random case. Although algebraic decay of branches is plausible for longer times, it appears unlikely for the relatively short time scales of branched flow, as discussed later.  In any case, besides the chaotic part of the flow, we must consider superwires, which are indefinitely stable branches associated with KAM islands in phase space. In random potentials, these islands occupy a negligible portion of phase space and do not significantly affect trajectory flow. However, in periodic potentials, the KAM islands fill a considerable amount of phase space and therefore carry an appreciable fraction of the flow. This means that Eq.~\ref{eq:expdie} must be modified to take superwires into account, so for \textit{periodic} potentials we have
\begin{equation}\label{eq:expdieperiodic}
    N_b \propto e^{-\Omega t}+C,
\end{equation}
where the constant $C$ is related to the fraction of phase space occupied by the superwires.  

\begin{figure}
    \centering
    \includegraphics[width=0.5\textwidth]{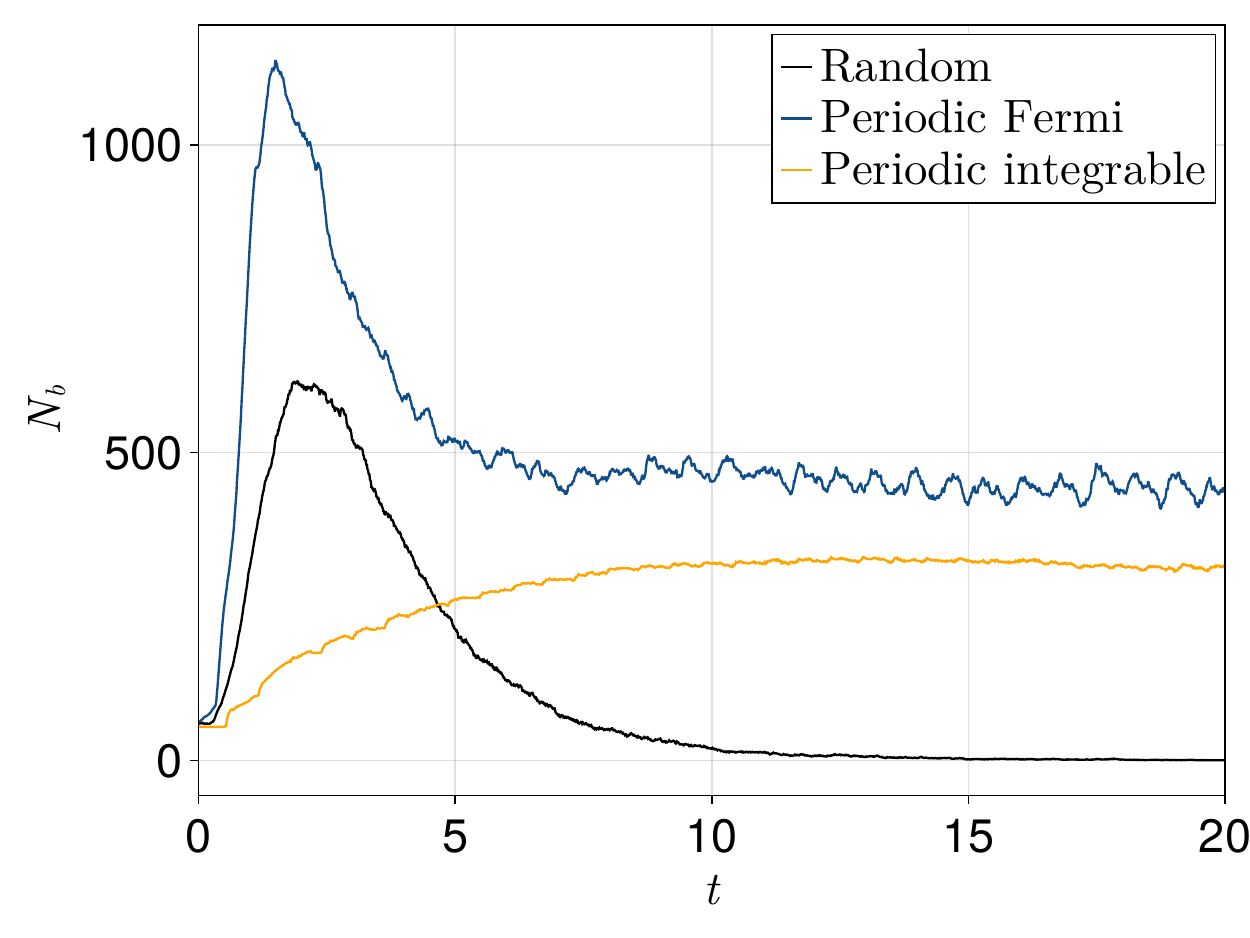}
    \caption{Number of branches over time, measured by the manifold tracking method (see Sec.~\ref{sec:counting}). After the initial growth, the number of branches decays exponentially. For the periodic potential of Eq.~(\ref{eq:sinV}), we observe that the number of branches does not go to zero, instead it reaches a horizontal asymptote. This phenomenon is attributed to the presence of superwires, which persist indefinitely. The plots have been averaged for 40 different potential orientations from 0 to $\pi/4$ with the entire manifold considered as it expands.}

    \label{fig:ExpDeath}
\end{figure} 

We have carried out numerical experiments to verify the previous theoretical arguments. First, we have computed the evolution of the number of branches $N_b$ as a function of the longitudinal direction $x$ for the three potentials proposed in Sec.~\ref{sec:potentials}. The results are depicted in Fig.~\ref{fig:ExpDeath}. In the random potential, the flow tends to a uniform ray density, and the number of branches $N_{b}$ drops to zero exponentially, as expected. 

However, in the case of the periodic integrable potential, we observe that the number of branches stabilizes after a transient period. Due to the presence of superwires associated with stable motion, the number of branches does not tend to zero as time progresses to infinity; instead, it converges to a nonzero value $C$, as predicted by Eq.~(\ref{eq:expdieperiodic}). The Fermi periodic potential exhibits mixed behavior, displaying both exponential decay and a constant floor $N_{b} \rightarrow C$ after the transient. To better understand the interaction between chaotic and periodic dynamics in non-integrable periodic potentials, we focus on the Fermi potential in this last case.

\begin{figure}
    \centering
    \includegraphics[width=0.5\textwidth]{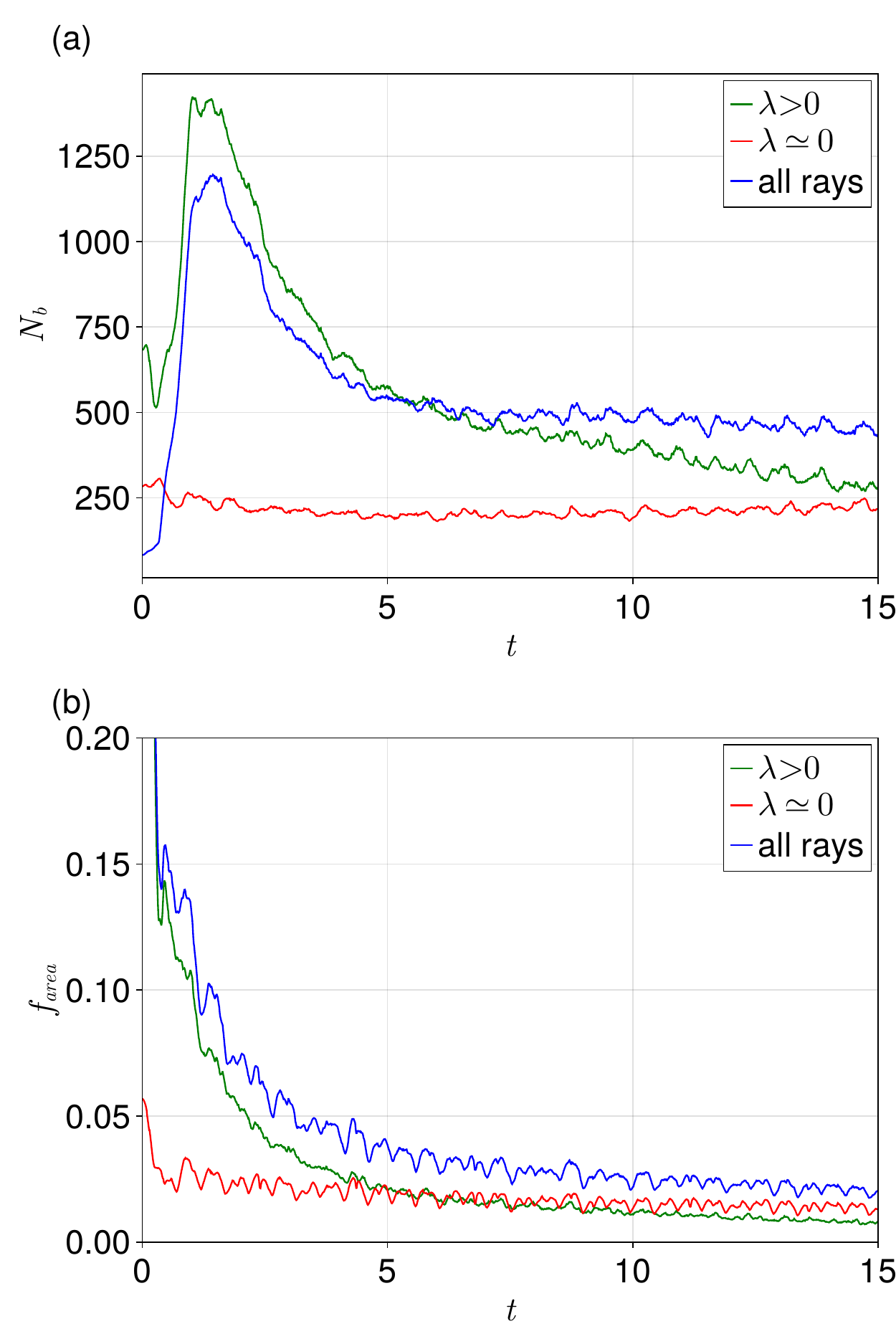}
    \caption{(a) Number of branches $N_b$ over time for the Fermi periodic potential. The branches have been classified depending on the stability of the rays measured with the maximum Lyapunov exponent $\lambda$. Only the part of the manifold within the bound of the original wavefront is taken into account, which explains the difference of scale of Fig.~\ref{fig:ExpDeath}. (b) Fraction of phase space occupied by the chaotic and stable flow, classified according to the Lyapunov exponents.}
\label{fig:fermideath}
\end{figure}

 In Fig.~\ref{fig:fermideath}-(a), the number of branches in the periodic Fermi potential is separated according to the Lyapunov exponents $\lambda$ of the trajectories. The number of total branches is the sum of contributions from the stable ($\lambda \approx 0$) and chaotic rays ($\lambda > 0$). The comparison with Fig.~\ref{fig:ExpDeath} reveals that the curve corresponding to the chaotic trajectories matches the behavior of the random potential, while the periodic part of the flow tends to a constant number of superwires. 

In our preceding theoretical analysis, we referred to the fraction of phase space occupied by the branches as the appropriate interpretation of the constant $C$ in Eq.~(\ref{eq:expdieperiodic}). Due to the potential overlap of multiple branches in phase space, a histogram-based approach is necessary to quantify how the branches populate the transverse space. Specifically, we identify bins where branches appear within the histogram and calculate the ratio $f_{area}$ as the number of bins with branches divided by the total number of bins examined over a fixed transverse interval. As depicted in Fig.~\ref{fig:fermideath}-(b), we observe a clear exponential decay in the chaotic dynamics contribution, while the superwires quickly stabilize around a floor value $C$.

Some results in existing literature suggest that the decay of the branches could be algebraic instead of exponential, due to the stickiness of the KAM islands~\cite{lai1992algebraic, motter2001dissipative}. The diffusion of trajectories in a Hamiltonian periodic potential similar to Eq.~\ref{eq:sinV} has been studied in~\cite{geisel1987}. The reported algebraic scaling law for the diffusion is directly related to the structure of the phase space, where nested cantori exist. While this may hold true for extended periods, our simulations have not yet demonstrated clear indications of such behavior. We conjecture that the branched flow regime occurs before stickiness significantly influences the dynamics.

\section{\label{sec:discussion}Discussion}
This work helps to clarify the relation between periodic potentials and branched flow. First, we shed some light into the definition of branched flow. In many papers, branched flow is considered to be a phenomenon restricted just to the so-called random potentials. Nevertheless, as far as for the growth of the number of branches is concerned, the behavior in random and non-integrable periodic potentials is exactly the same. This underscores the importance of the chaotic dynamics for the occurrence of branched flow, rather than the spatial disorder of the potential. 

Nonetheless, branched flow in periodic potentials also introduces intriguing novelties, particularly concerning superwires. The potential for stable motion in periodic systems engenders fresh behaviors, exemplified by the manner in which branches diminish. While branches associated with chaotic dynamics exhibit behaviors akin to those in random potentials, superwires persistently concentrate flow, creating a plateau in the decay of the branches. Superwires possess significant inherent interest \cite{graf2024chaos}, and their application across various fields holds a promising potential. Nonetheless, it is conceivable to envision a periodic potential with minimal KAM islands in phase space, producing branch behaviors practically indistinguishable from those in random potentials, thus underscoring the pivotal role of chaotic dynamics in branched flow.

We hope our work will stimulate further research into branched flow in periodic potentials, an area ripe with theoretical and practical connections waiting to be explored.

\section*{Acknowledgments}

This work has been supported by the Spanish State Research Agency (AEI), the European Regional Development Fund (ERDF, EU) under Project No.~PID2019-105554GB-I00, and the Research Council of Finland under Project No. 349956 (ManyBody2D). A.D. thanks Prof. Eric J. Heller for introducing him to branched flow and for his continued mentoring.

\bibliography{biblio_branched}

\end{document}